\documentclass{article}
\usepackage{authblk}

\usepackage{graphicx}
\usepackage{amssymb}
\usepackage{pifont}

\begin{document}

\title{Microdosimetry spectra and RBE of $^1$H, $^4$He, $^7$Li and $^{12}$C nuclei in water studied with Geant4}

\author[1,2]{Lucas~Burigo%
  \thanks{Electronic address: \texttt{burigo@fias.uni-frankfurt.de}; Corresponding author}}
\author[1,3]{Igor~Pshenichnov%
  \thanks{Electronic address: \texttt{pshenich@fias.uni-frankfurt.de}}}
\author[1,4]{Igor~Mishustin}
\author[1,2]{Marcus~Bleicher}

\affil[1]{Frankfurt Institute for Advanced Studies, Johann Wolfgang Goethe University, 60438 Frankfurt am Main, Germany}
\affil[2]{Institut f\"ur Theoretische Physik, Johann Wolfgang Goethe University, 60438 Frankfurt am Main, Germany}
\affil[3]{Institute for Nuclear Research, Russian Academy of Sciences, 117312 Moscow, Russia}
\affil[4]{Kurchatov Institute, Russian Research Center, 123182 Moscow, Russia}

\maketitle

\begin{abstract}
A Geant4-based Monte Carlo model for Heavy-Ion Therapy (MCHIT) is used to study radiation fields of $^1$H, $^4$He, $^7$Li and $^{12}$C beams with similar ranges ($\sim$160-180~mm) in water. Microdosimetry spectra are simulated for wall-less and walled Tissue Equivalent Proportional Counters (TEPCs) placed outside or inside a phantom, as in experiments performed, respectively, at NIRS, Japan and GSI, Germany. The impact of fragmentation reactions on microdosimetry spectra is investigated for $^4$He, $^7$Li and $^{12}$C, and contributions from nuclear fragments of different charge are evaluated for various TEPC positions in the phantom. The microdosimetry spectra measured on the beam axis are well described by MCHIT, in particular, in the vicinity of the Bragg peak. However, the simulated spectra for the walled TEPC far from the beam axis are underestimated. Relative Biological Effectiveness (RBE) of the considered beams is estimated using a modified microdosimetric-kinetic model. Calculations show a similar rise of the RBE up to 2.2--2.9 close to the Bragg peak for helium, lithium and carbon beams compared to the modest values of 1--1.2 at the plateau region. Our results suggest that helium and lithium beams are also promising options for cancer therapy.
\end{abstract}

\section{Introduction}

Presently proton and $^{12}$C beams are successfully used for cancer treatment~\cite{Schardt2010,Durante2010,Jensen2011,Kamada2012}. Other projectiles, e.g. $^4$He and $^7$Li, may  differ in their biological action from $^{12}$C nuclei, but still have beam divergence similar to $^{12}$C, and thus can be considered as new treatment options~\cite{Kempe2007}. Beams of protons, helium, lithium, beryllium, carbon, and neon nuclei were recently compared~\cite{Remmes2012} from the point of view of their advantage to spare healthy tissues with respect to radiobiological parameters ($\alpha/\beta$ ratio) of normal and target tissues. Other authors~\cite{Kempe2007,Kantemiris2011} studied the depth-dose and linear energy transfer (LET) distributions  of protons, $^4$He, $^7$Li and $^{12}$C in water using the Monte Carlo codes SHIELD-HIT and FLUKA, respectively.

In view of possible applications of nuclei other than carbon in cancer therapy, the quality of radiation fields created by such projectiles has to be studied. For this purpose we have used our Monte Carlo model for Heavy-Ion Therapy (MCHIT)~\cite{Pshenichnov2006,Pshenichnov2007,Pshenichnov2008,Pshenichnov2010,Mishustin2010}  based on the Geant4 toolkit~\cite{Agostinelli2003,Allison2006}. In a recent publication~\cite{Pshenichnov2008} we have compared the depth-dose distributions for various projectiles propagating in water. These calculations took into account the fragmentation of projectile nuclei in collisions with nuclei of the medium. The calculated dose profiles were compared with experimental data where available. In particular, the depth-dose profiles for $^3$He nuclei in water were studied along with the distributions of positron-emitting nuclei produced by these projectiles~\cite{Pshenichnov2007}.

While the capabilities of the Geant4 toolkit to model propagation of protons and carbon nuclei in tissue-like media were already demonstrated in several publications, see e.g.~\cite{Moteabbed2012,Seravalli2012} and~\cite{Lechner2010}, much less attention was paid to simulations with other projectiles, e.g. $^4$He and $^7$Li. One may expect that due to a reduced total reaction cross section of these light projectiles, the importance of fragmentation reactions on the corresponding dose distribution will be reduced with respect to $^{12}$C. On the other hand one can note, that while boron or beryllium nuclei are frequently produced by $^{12}$C with their $Z^2$ close to the projectile nucleus, $^4$He usually fragments into a proton, a neutron and a deuteron resulting in a rapid drop of $Z^2$. This indicates that in addition to the known reduction of the total fragmentation cross section with the decrease of the projectile mass, the composition of secondary fragments has also to be taken into account. In turn, this will lead to different biological properties of such beams.

As demonstrated recently, MCHIT describes well microdosimetry spectra for neutron and carbon-ion beams~\cite{Burigo2013}. In this work we present Monte Carlo calculations of microdosimetry distributions for proton, $^4$He, $^7$Li and $^{12}$C beams in water and compare results with experimental data.
The obtained microdosimetry spectra are used to estimate the Relative Biological Effectiveness (RBE) of these nuclei both on the beam axis and away from it. Differences in the physical and biological properties  of these therapeutic beams are discussed.

\section{Materials and Methods}

\subsection{Microdosimetric Measurements}

Patterns of energy deposition in tissue by ionizing particles at the micrometer scale can be measured by Tissue Equivalent Proportional Counters (TEPC). The amount of energy delivered to the TEPC sensitive volume by particles traversing  the detector fluctuates due to the stochastic nature of particle transport in media~\cite{ICRU1983}. Therefore, the lineal energy $y=\epsilon/\bar{l}$, where $\epsilon$ is the deposited energy in a given event and $\bar{l}$ is the mean chord length of the sensitive volume, changes from one event to another and a probability density $f(y)$ can be measured. This probability density is usually characterized by the frequency-mean lineal energy, $\bar{y}_f$, and dose-mean lineal energy, $\bar{y}_d$, defined~\cite{ICRU1983} as
\begin{equation}
\bar{y}_f = \int_{0}^{\infty} yf(y){\rm d}y \ ,
\label{eq:y_f_definition}
\end{equation}
\begin{equation}
\bar{y}_d = \frac{1}{\bar{y}_f}\int_{0}^{\infty} y^2 f(y) {\rm d}y = \int_{0}^{\infty} yd(y){\rm d}y  .
\label{eq:y_d_definition}
\end{equation}
Here $d(y)\equiv yf(y)/\bar{y}_f$ is introduced as the dose probability density.
The saturation-corrected dose-mean lineal energy, $y^*$, defined~\cite{ICRU1983} as:
\begin{equation}
y^* = \frac{ y_0^2 \int_{0}^{\infty} \left( 1-\exp{\left(-y^2/y_0^2\right)}\right) f(y){\rm d}y }{ \int_{0}^{\infty} yf(y) {\rm d} y } \ ,
\label{eq:ystar}
\end{equation}
is introduced to account for the saturation of biological effects induced by high-LET radiation. 
The saturation parameter $y_0$ = 150~keV/$\mu$m~\cite{Kase2006} is used in our study.

\subsubsection{Microdosimetry of $^1$H and $^4$He beams}

\begin{figure}[htb]
\centering
\includegraphics[width=0.6\columnwidth]{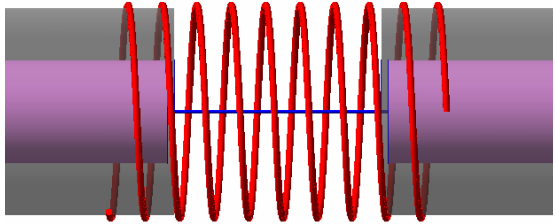}
\caption{Inner geometry of a wall-less TEPC implemented in MCHIT.}
\label{fig:walllessTEPC}
\end{figure}

Microdosimetry measurements for 160~MeV proton and 150$A$~MeV helium beams were performed at the Heavy Ion Medical Accelerator in Chiba (HIMAC) of the National Institute of Radiological Science (NIRS), Japan~\cite{Tsuda2012}. The Bragg peaks in water for such proton and helium projectiles are located at the depth of $\sim 175.6$ and $\sim 158$~mm, respectively. A wall-less TEPC simulating a tissue volume of 0.72~$\mu$m in diameter was employed along with a range shifter for energy degradation. The geometry of the wall-less TEPC including the anode, cathode, insulators, field tubes and beam window components, is implemented in MCHIT, as illustrated in Fig.~\ref{fig:walllessTEPC}. The helical geometry of the cathode is simulated by small G4Torus segments displaced along and rotated around the anode wire. A total of 360 torus segments per pitch is used.
The range shifter is simulated as a water layer in front of the detector device. Microdosimetry spectra at various beam energies were measured at HIMAC by changing the thickness of the range shifter. The uncertainties in the water equivalent thickness of the range shifter and beam elements in the experimental set-up were not reported. Taking into account that the precise dimension and material properties of the range shifter were not reported, and also that the calculation of water-equivalent thickness is influenced by uncertainties in the stopping power of ions, in our simulations we decided to adjust the thickness of the range shifter in order to reproduce the position of the main peak in microdosimetry spectra which is associated with primary beam particles traversing the detector. The water equivalent thickness used in the simulations are smaller than the estimated value for the experimental set-up by about 3\% and 0.8\% for proton and helium beams, respectively. A total of $4~\times~10^7$ events were simulated for each microdosimetry spectrum. The experimental data were reported as $yf(y)/\bar{y_f}$ distributions.

\subsubsection{Microdosimetry of $^7$Li and $^{12}$C beams}

\begin{figure}[htb]
\centering
\includegraphics[width=0.5\columnwidth]{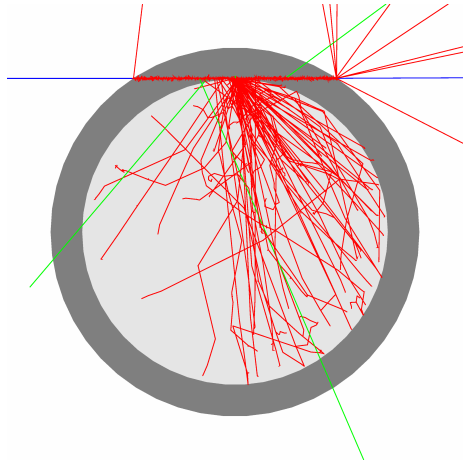}
\caption{Geometry of a walled TEPC implemented in MCHIT with tracks of secondary particles produced in one event for a $^{12}$C nucleus (blue track) crossing the TEPC wall. Red and green tracks represent secondary electrons and photons, respectively. The external aluminium cap is not shown.}
\label{fig:walledTEPC}
\end{figure}

Microdosimetry spectra for 185$A$~MeV $^7$Li and 300$A$~MeV $^{12}$C beams were measured  at GSI,  Germany~\cite{Martino2010} at several positions inside a water phantom. A compact walled TEPC was employed to measure microdosimetry spectra for a tissue equivalent volume of 2.7~$\mu$m in diameter.
The total water equivalent thickness of the PMMA wall of the water phantom and beam-line elements used in experiment~\cite{Martino2010} amounts to 25.1~mm.
Similar to the calculations for $^1$H and $^4$He beams described above, the TEPC positions were adjusted in order to reproduce the position of the main peak in the spectra when the TEPCs were placed on the axis in the vicinity of the Bragg peak of $^7$Li and $^{12}$C beams. The shift of the TEPC for the carbon beam (2~mm deeper position which corresponds to 1\% of the range of carbon ion) was discussed elsewhere~\cite{Burigo2013}. As for $^7$Li beam, the microdosimetry spectrum measured for the TEPC located at the Bragg peak could be only reproduced by calculations if the TEPC is further shifted to deeper position by 6.7~mm. One must notice that our simulations have shown that the range in the water phantom for 185$A$~MeV $^7$Li beam is 5~mm deeper than the one for the 300$A$~MeV $^{12}$C beam while it was reported that the specific energies for the carbon and lithium beams in the experimental set-up were chosen such that the residual range in the water phantom is the same for both ions~\cite{Martino2010}. The simulated geometry of the walled TEPC, see Fig.~\ref{fig:walledTEPC}, includes its external aluminium cap, a 1.27~mm thick wall made of Shonka A-150 tissue-equivalent plastic and an inner spherical gas cavity. The cathode and anode elements in the detection region were not simulated. A total of $10^7$ histories of primary ions and all their secondary particles were simulated for each microdosimetry spectrum. Each walled TEPC detector placed far from the beam axis was represented in simulations by a number of virtual TEPCs placed on a ring at the same depth~\cite{Burigo2013}. Due to this special arrangement of virtual TEPC detectors, the counting rate of events per primary track is significantly increased in each physical TEPC. The experimental data were reported as $yd(y)$ distributions normalized to the number of incoming primary ions.

\subsection{Monte Carlo modelling by MCHIT}\label{sec:physical_models}

Microdosimetry spectra and related variables are calculated by the Monte Carlo method using MCHIT. This code is a Geant4 application based on the version 9.5 with patch 02 of this toolkit.
Description of the physics models included in the Geant4 toolkit is given elsewhere~\cite{Ivanchenko2011}.
A set of models which are relevant to a particular problem should be activated by the application developer. In MCHIT we make use of the so-called predefined physics lists along with customized physics lists. 
The predefined physics lists are provided by Geant4 developers and distributed along with Geant4. Separate physics lists for electromagnetic and hadronic physics are kept for convenience. More details on the physics models used in MCHIT can be found elsewhere~\cite{Burigo2013}.

\subsubsection{Electromagnetic physics}

Three different predefined physics list for electromagnetic processes are employed in the following calculations, namely G4EmStd (which uses the ``Standard Electromagnetic Physics Option 3''), G4EmPen (which uses the Penelope models for low energy processes) and G4DNA (which uses the track structure models known as G4DNA models). The involved physics models simulate the energy loss and straggling of primary and secondary charged particles due to interaction with atomic electrons. G4EmStd and G4EmPen are based on continuous slowing-down approximation and algorithms of multiple Coulomb scattering of charged particles on atomic nuclei. G4DNA explicitly simulate each single electromagnetic interaction.  These physics lists  differ in the capability of models to produce and transport low-energy $\delta$-electrons.  The low-energy thresholds for production of $\delta$-electrons are 990~eV for G4EmStd and 100~eV for G4EmPen, while all $\delta$-electrons are produced and transported by G4DNA. A customized physics list, G4EmPen+IonGas, which is based on G4EmPen and the models describing the ionization of gas media by ions, is also used in calculations.

\subsubsection{Hadronic physics}

A customized physics list was implemented for the description of hadronic processes. The first fast stage of nucleus-nucleus collisions is described by two Geant4 models, the Light Ion Binary Cascade model (G4BIC)~\cite{Folger2004} for proton, helium and lithium beams, and the Quantum Molecular Dynamics model (G4QMD)~\cite{Koi2010} for carbon beams. As a result of simulation of a nucleus-nucleus collision some excited nuclear fragments are produced in addition to free nucleons. Therefore, the G4ExcitationHandler of Geant4 is used to simulate subsequent decays of excited nuclear fragments by applying various de-excitation models depending on the mass and excitation energy of these fragments. The Fermi break-up model (G4FermiBreakUp) is applied to excited nuclei lighter than fluorine. It is designed to describe explosive disintegration of excited light nuclei~\cite{Bondorf1995} and it is highly relevant to collisions of light nuclei with nuclei of 
tissue-like materials.
For heavier excited nuclei either the evaporation model~\cite{Weisskopf1940} can be used at low excitations (below 3~MeV per nucleon), or the statistical multifragmentation model  (SMM)~\cite{Bondorf1995} at higher excitation energies. Generally, the inclusion of fragmentation reactions helps to describe the yield of intermediate-mass fragments.

\subsection{Theoretical estimation of RBE}\label{sec:RBE}

The RBE for a given kind of mammalian cells irradiated by ions can be estimated using the microdosimetric-kinetic (MK) model~\cite{Hawkins1996,Hawkins2003}. The MK model combines a microdosimetric description of energy deposition events in sub-nuclear domains of irradiated cells with a kinetic description of creation and repair 
of radiation-induced lesions. In this phenomenological model the linear-quadratic relation for the cell survival curve is derived for low-LET radiation by assuming a Poisson distribution of lethal lesions in the domain volume~\cite{Hawkins1994}. In the case of high-LET particles a non-Poisson distribution is applied as a correction for the overkill effect~\cite{Hawkins2003}. An alternative approach was also proposed~\cite{Kase2006}, which exploits the dose-mean lineal energy corrected for saturation. This modified version of the MK model~\cite{Kase2006} allows to use microdosimetric spectra measured by TEPCs to estimate survival rates of irradiated cells. 

In particular, the surviving fraction $S$ of human salivary gland (HSG) tumour cells is expressed as exponent of 
a linear-quadratic function of the dose $D$,
\begin{equation}
S = \exp{\left[ -\alpha D -\beta D^2 \right]} \ ,
\end{equation}
where
\begin{equation}
\alpha = \alpha_0 + \frac{\beta}{\rho \pi r_d^2} y^* \ ,
\label{eq:alpha}
\end{equation}
with the following model parameters: $\alpha_0$ = 0.13~Gy$^{-1}$ as a constant that represents the initial slope of the survival fraction curve in the limit of zero LET, $\beta$ = 0.05~Gy$^{-2}$ as a constant independent of LET, $\rho = $1~g/cm$^3$ as the density of tissue and $r_d = $0.42~$\mu$m as the radius of a sub-cellular domain in the MK model. Equation~(\ref{eq:alpha}) establishes a relation between $y^*$ and the $\alpha$-parameter of the linear-quadratic model irrespective of the specific ion species. This relation reflects the fact that an excessive local energy deposition is inefficient to boost a given biological effect~\cite{ICRU1983}. This leads to a reduction of the RBE known as the saturation effect. 

According to the modified MK model the RBE for HSG cells can be estimated using the following relation~\cite{Kase2011}:
\begin{equation}
RBE_{10} = \frac{2\beta D_{10,R}}{\sqrt{\alpha^2-4\beta\ln{\left(0.1\right)}}-\alpha},
\label{eq:RBE}
\end{equation}
where $D_{10,R}$ = 5.0~Gy is the 10\% survival dose of the reference radiation (200~kVp X-rays) for HSG cells~\cite{Kase2006}.

As shown in~\cite{Kase2006,Sato2011}, the microdosimetric parameter $y^*$ is well suited to estimate the $\alpha$-parameter for HSG and other cell lines for a large variety of projectiles including proton, helium and carbon nuclei. Therefore, the RBE values for various therapeutic beams can be obtained from the corresponding 
microdosimetry spectra. In the present study the aforementioned modified MK model is used to estimate the RBE of proton,  $^4$He, $^7$Li and $^{12}$C beams for HSG cells. Firstly, microdosimetric spectra are calculated by means of the MCHIT model placing the walled TEPC described above behind a range shifter made of water. The thickness of the range shifter was varied in order to calculate the microdosimetry spectra at different water-equivalent depths. The pressure of the TE gas was set to simulate a tissue sphere of 1~$\mu$m in diameter and the detector was irradiated by a broad beam. This set-up mimics the experimental conditions used by Kase et al. in measurements of $y^*$ for a 290$A$~MeV $^{12}$C beam~\cite{Kase2006}. RBE$_{10}$ values for HSG cells irradiated by carbon beams were estimated according to Eq.~(\ref{eq:RBE}) with the experimental and simulated values of $y^*$ and used for the MCHIT validation. RBE$_{10}$ for proton, helium and lithium were also calculated for the sake of comparison. In order to evaluate the biological effectiveness away from the beam axis, a second set-up was devised. In this case the TEPC was placed inside a water phantom at depths close to the Bragg peak and 2~cm away from the axis of a pencil-like beam.
A large number ($\sim 10^7$) of beam particle histories were simulated to ensure that statistical fluctuations of the calculated $y^*$ are negligible in all set-up configurations.

\section{Results and Discussion}\label{sec:results}

\subsection{Contribution of secondary fragments to the microdosimetry spectra}

\begin{figure}[htb]
\centering
\includegraphics[width=0.6\columnwidth]{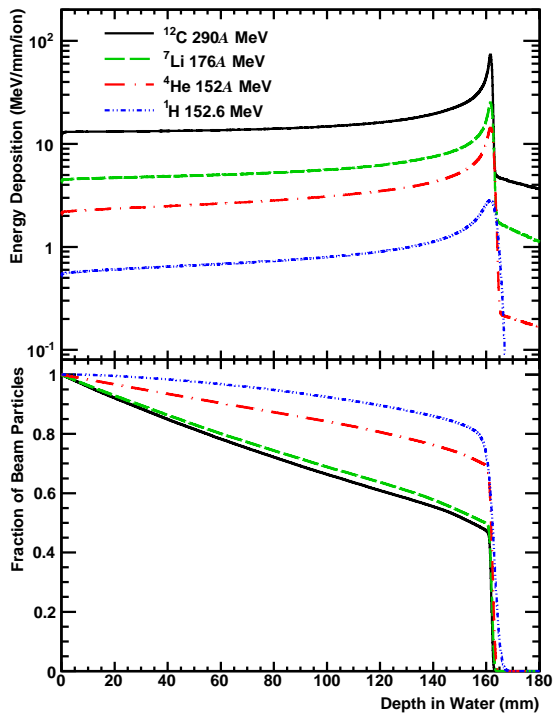}
\caption{Calculated depth-dose distributions in water for $^1$H, $^4$He, $^7$Li and $^{12}$C beams considered in the present work (top panel) and the attenuation factor for these beams due to nuclear reactions (bottom panel).}
\label{fig:depth_profiles}
\end{figure}

The role of nuclear reactions to attenuate the intensity of $^1$H, $^4$He, $^7$Li and $^{12}$C beam particles while they propagate in water can be well understood from Fig.~\ref{fig:depth_profiles}.
The energy per projectile nucleon was taken as 152.6~MeV for $^1$H, 152~MeV for $^4$He, 176~MeV for
$^7$Li and 290~MeV for $^{12}$C. With this choice of energies all the beams have the Bragg peaks at 161.6~mm depth in water. In Fig.~\ref{fig:depth_profiles} the fractions of surviving beam nuclei at certain depth (bottom) are plotted together with the corresponding depth-dose curves (top). 
As seen from Fig.~\ref{fig:depth_profiles}, $\sim 50$\% of $^7$Li and $^{12}$C  beam nuclei are lost before they reach the depth of the Bragg peak, where they finally stop. Nuclear reactions are less frequent for $^1$H and $^4$He beams, as only $\sim 20$\% of protons and $\sim 30$\% of alphas participate in nuclear reactions before they stop.

In the experiments of Tsuda et al.~\cite{Tsuda2012} the $^1$H and $^4$He projectiles entered the TEPC after traversing 150--170~mm of water when their energies were reduced to 17--38 MeV per nucleon. As seen from Fig.~\ref{fig:depth_profiles}, this corresponds to a TEPC placed close to the Bragg peak. Even at this deep location $\sim 70$--80\% of beam protons and alphas reached the TEPC without participating in nuclear reactions. Therefore, a relatively small contribution of secondary fragments to microdosimetry spectra is expected for $^1$H and $^4$He beams. Nevertheless, as demonstrated below, for $^4$He beam the contributions of specific secondary fragments to microdosimetry spectra can still be identified. The contributions of secondary fragments to microdosimetry spectra obtained with $^7$Li and $^{12}$C~\cite{Martino2010}  are expected to be much larger as compared with $^1$H and $^4$He beams. Indeed, as can be estimated from Fig.~\ref{fig:depth_profiles}, a noticeable number of secondary fragments of $^7$Li and $^{12}$C traverse TEPCs placed at the depth of $\sim 50$~mm and especially near the Bragg peak in agreement with the measurements by Martino et al.

The LETs of protons and ions vary significantly in the range of the kinetic energies used for radiation therapy. The lineal energy $y$, which is measured by TEPC, serves as an estimation of LET and its frequency distribution includes contributions from various particles traversing the detector. However, since different particles may contribute to similar or overlapping domains of $y$, such contributions can not be easily disentangled in experiment~\cite{Waker1995}, unless a complicated procedure to identify the charges of fragments is involved. Alternatively, specific contributions from certain particles can be scored and identified in calculations using the Monte Carlo method. Once the validity of Monte Carlo modelling is confirmed by a good agreement of calculated and measured spectrum, the contributions from specific particles can be reliably identified. The study of microdosimetry spectra collected far from the beam helps to evaluate the accuracy of nuclear fragmentation models since such spectra are built entirely by secondary fragments. At the same time, the Monte Carlo modelling of microdosimetry spectra opens the possibility to understand the quality of radiation in mixed radiation fields.

\subsection{Beam of $^1$H in water}

\begin{figure}[htb]
\begin{minipage}[t]{0.5\linewidth}
\includegraphics[width=1.0\columnwidth]{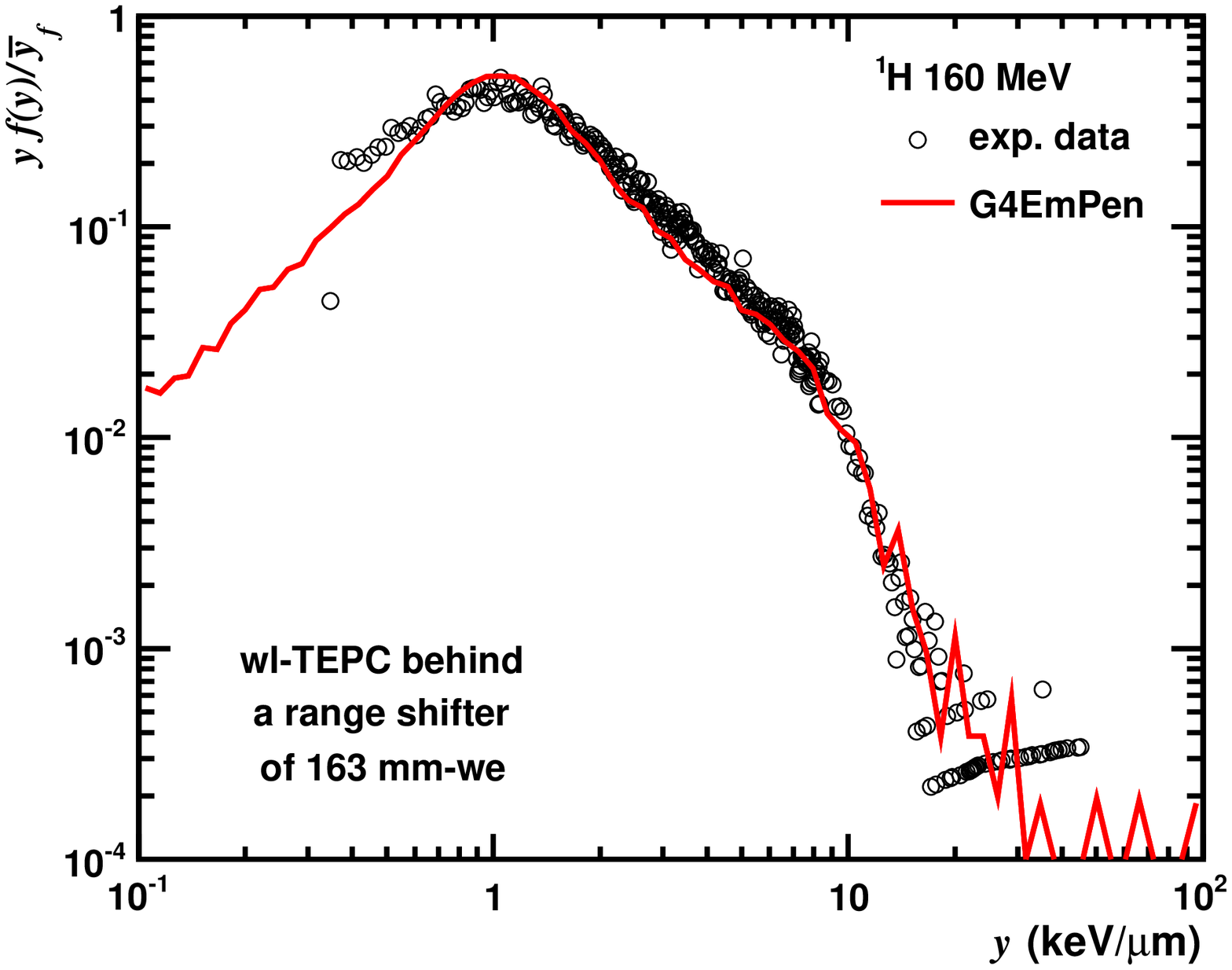}
\caption{Microdosimetry spectrum calculated with G4EmPen behind the range shifter (163~mm w.e.) for a 160 MeV proton beam. Circles represent experimental data~\protect\cite{Tsuda2012} corresponding to the estimated average proton energy of 38~MeV at the entrance to TEPC.}
\label{fig:spectra_H_38}
\end{minipage}
\hspace{0.01\linewidth}
\begin{minipage}[t]{0.5\linewidth}
\includegraphics[width=1.0\columnwidth]{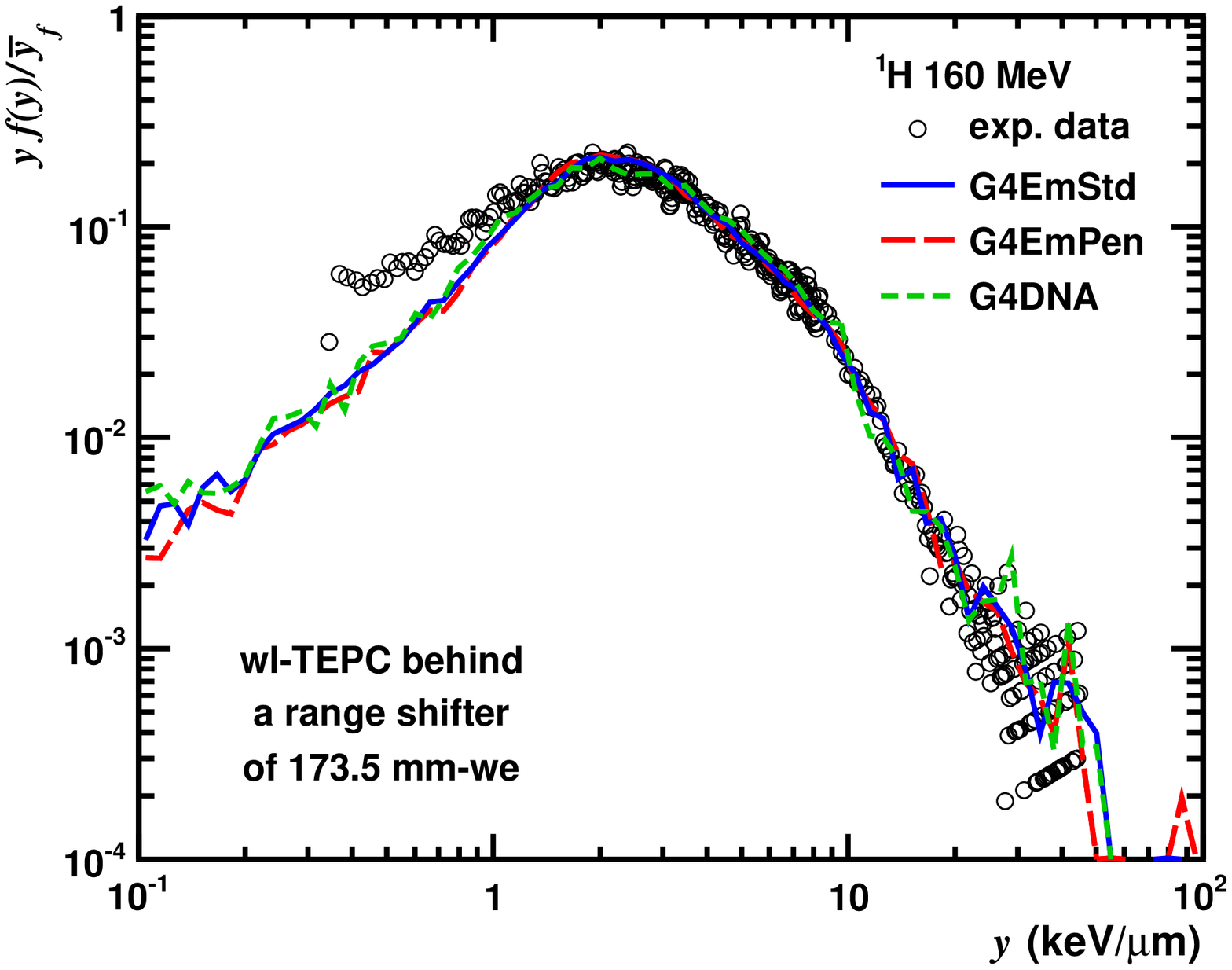}
\caption{Microdosimetry spectra calculated with G4EmPen, G4EmStd and G4DNA behind the range shifter (173.5~mm w.e.) for a 160 MeV proton beam. Circles represent experimental data~\protect\cite{Tsuda2012} corresponding to the estimated average proton energy of 20~MeV at the entrance to TEPC.}
\label{fig:spectra_H_20}
\end{minipage}
\end{figure}

The simulated microdosimetry spectrum for a proton beam traversing a $163$~mm range shifter made of water is shown in Fig.~\ref{fig:spectra_H_38} along with experimental data~\cite{Tsuda2012}. This spectrum corresponds to the TEPC position at $\sim 13$~mm before the Bragg peak. Similar results, but for a TEPC placed closer to the Bragg peak are shown in Fig.~\ref{fig:spectra_H_20}. In the latter case the TEPC is traversed by less energetic protons which have a higher LET. This is confirmed by the shift of the maximum of the microdosimetry spectra to larger $y$, which can be seen  by comparing Fig.~\ref{fig:spectra_H_20} and Fig.~\ref{fig:spectra_H_38}.
One can clearly see that the model systematically underestimates events with very low lineal energy ($y < 1$~keV/$\mu$m) in all simulations with the wall-less TEPC. In order to investigate whether this is related to the limitation for production of $\delta$-electrons with G4EmStd and G4EmPen, we carried out a simulation with G4DNA with results  presented in Fig.~\ref{fig:spectra_H_20}. The distribution of particle energies downstream of the TEPC window was first calculated with G4EmPen and then used as an input to simulations with G4DNA. The simulated gas cavity of the TEPC was replaced in G4DNA simulations by a homogeneous water-equivalent volume. As shown in Fig.~\ref{fig:spectra_H_20}, G4EmPen and G4DNA provide statistically equivalent results even though the simulation with G4EmPen accounts for a detailed geometric description of the detector, while G4DNA works only with an equivalent cylindrical water volume. Since there are no low-energy limits for production of $\delta$-electrons in simulations with G4DNA, one should conclude that the deficit of events with very low $y$ is not related to limitations of electron transport.

As seen from Fig.~\ref{fig:spectra_H_20}, the microdosimetric spectrum of protons calculated for the macroscopic-size TEPC agree well with the microdosimetric spectrum calculated for the equivalent microscopic volume of water with the G4DNA physics list. This confirms the basic assumption of the microdosimetry technique and justifies the calculation of microdosimetric spectra using the continuous slowing-down approximation.

\subsection{Beam of $^4$He in water}

\begin{figure}[htb]
\begin{minipage}[t]{0.5\linewidth}
\centering
\includegraphics[width=1.0\columnwidth]{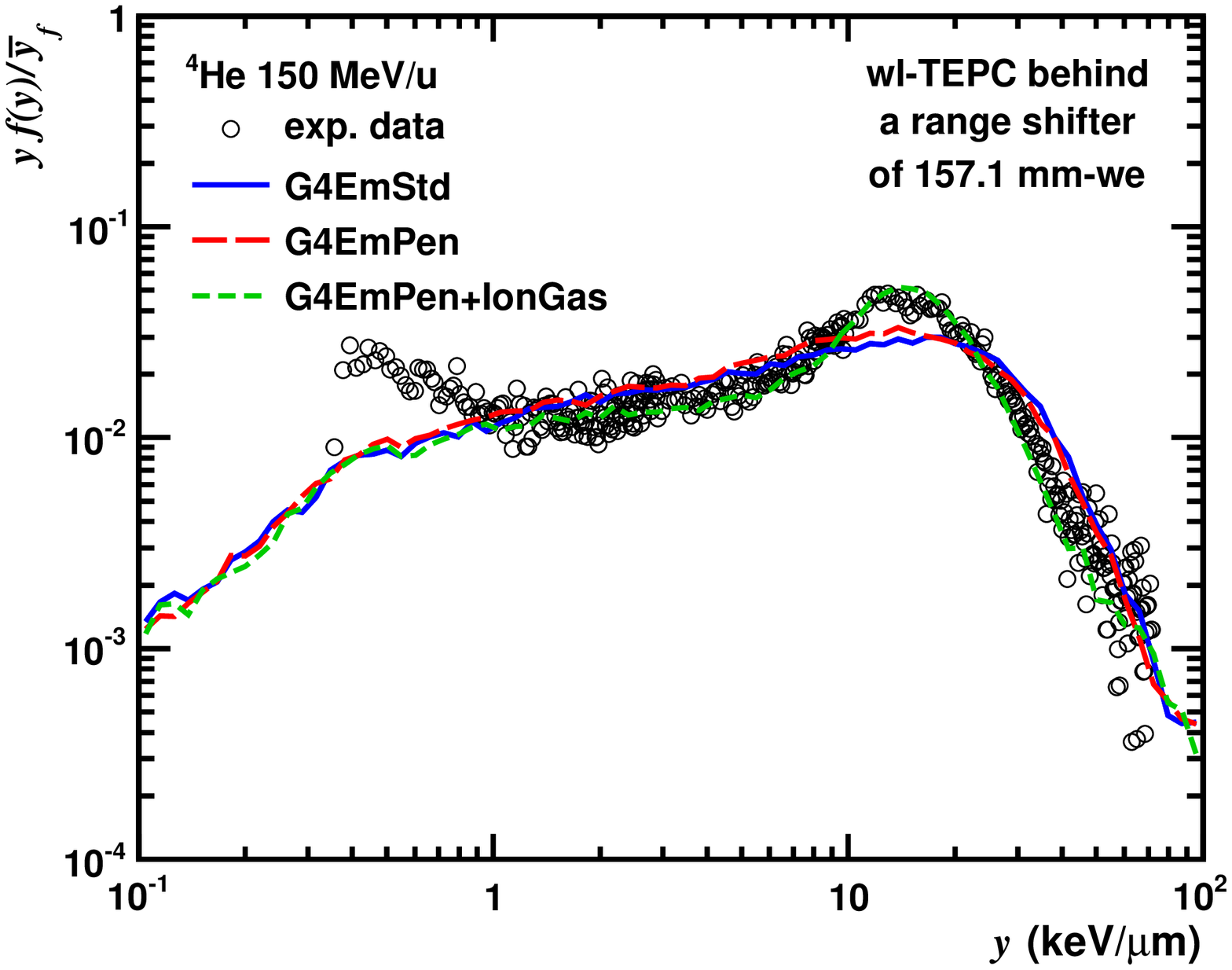}
\caption{Microdosimetry spectrum behind a range shifter (157.1~mm w.e.) for a 150$A$ MeV $^4$He beam calculated with G4EmStd, G4EmPen and G4EmPen+IonGas. Circles represent experimental data~\protect\cite{Tsuda2012} corresponding to the estimated average $^4$He energy of 17$A$~MeV at the entrance to TEPC.}
\label{fig:spectra_He_17_em}
\end{minipage}
\hspace{0.01\linewidth}
\begin{minipage}[t]{0.5\linewidth}
\centering
\includegraphics[width=1.0\columnwidth]{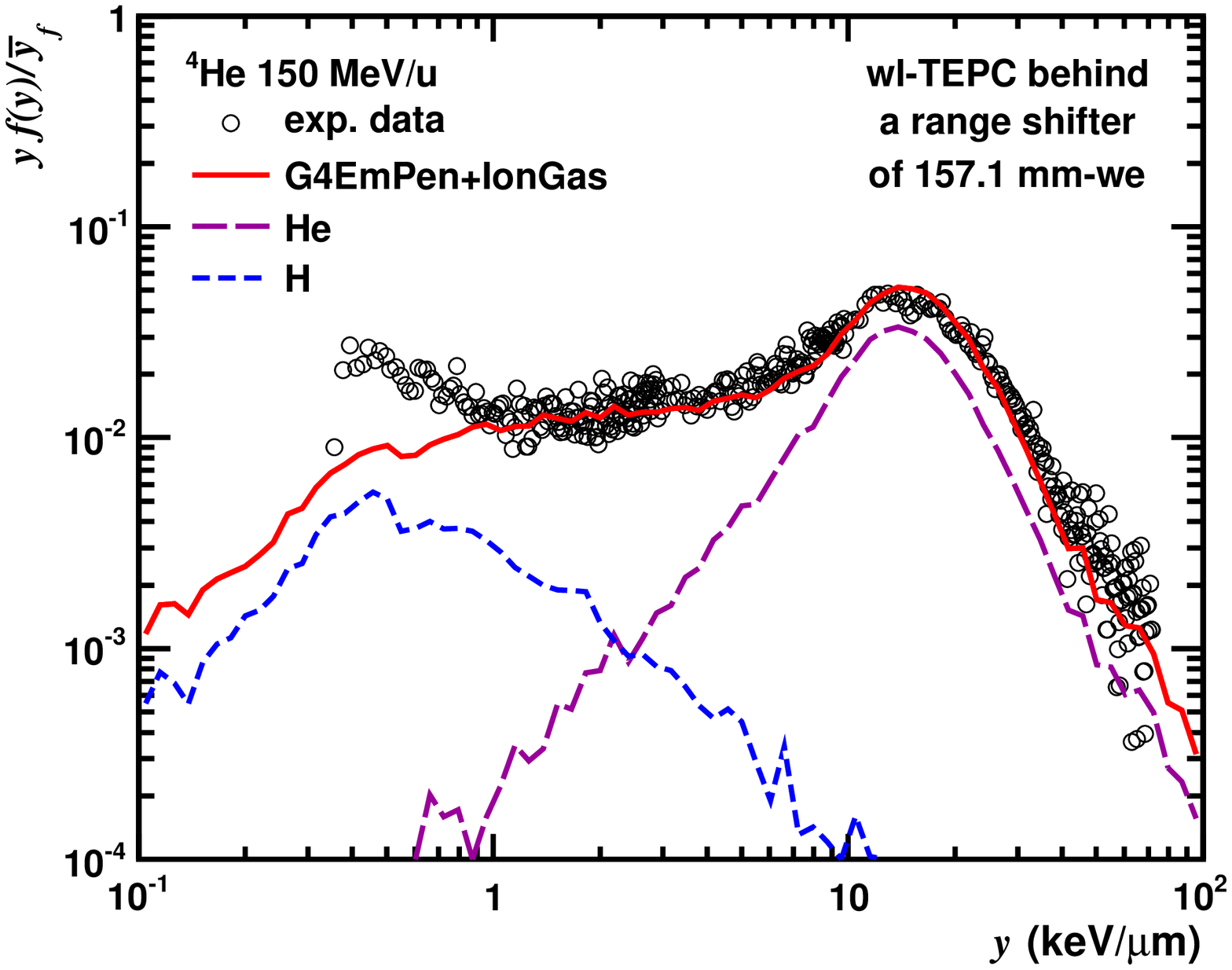}
\caption{Same as in Fig.~\protect\ref{fig:spectra_He_17_em}, but with microdosimetry spectrum calculated only with G4EmPen+IonGas. Specific contributions from hydrogen and helium nuclei are shown separately as explained in the figure legend.}
\label{fig:spectra_He_17}
\end{minipage}
\end{figure}

\begin{figure}[htb]
\begin{minipage}[t]{0.5\linewidth}
\centering
\includegraphics[width=1.0\columnwidth]{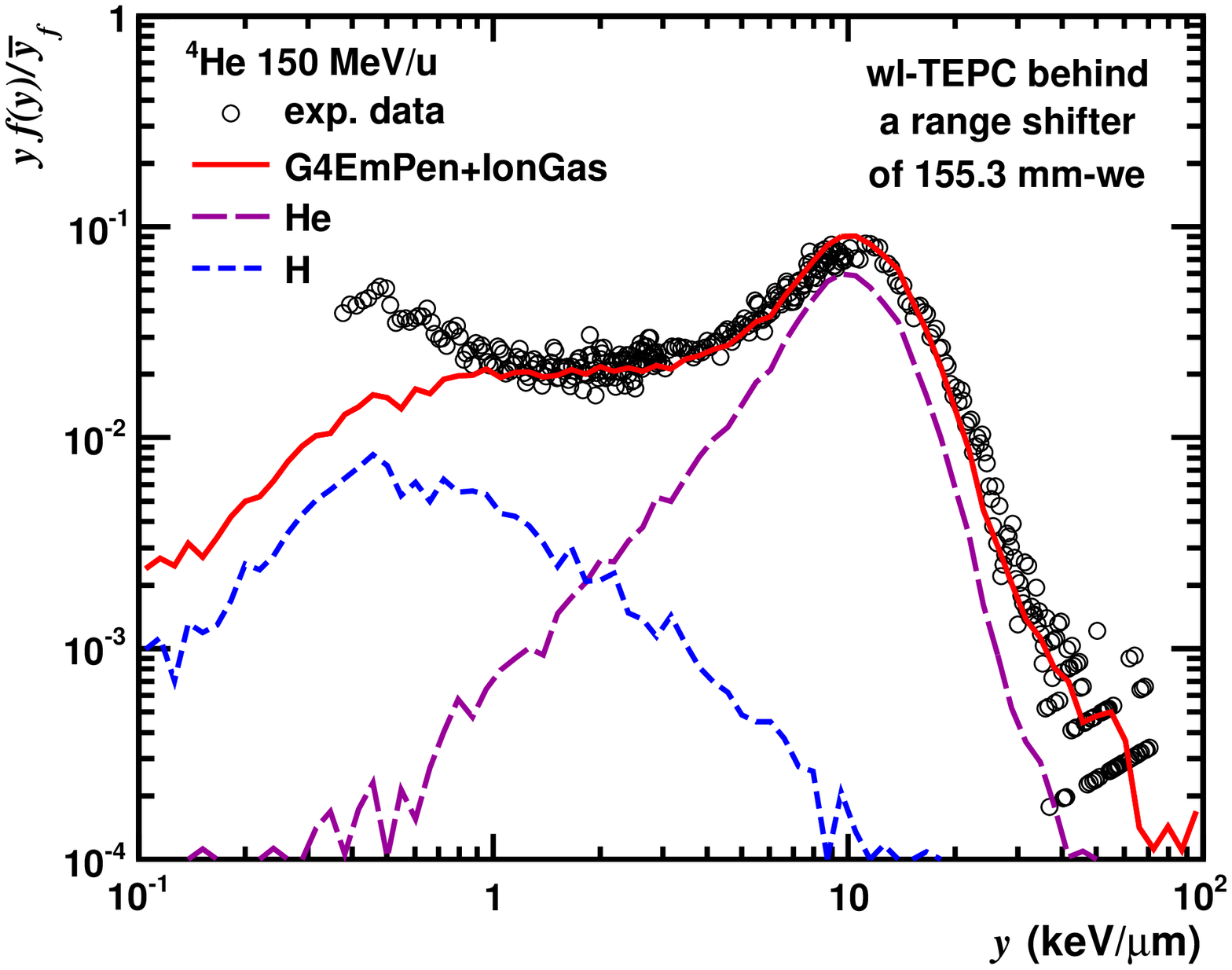}
\caption{\protect\label{fig:spectra_He_22}Microdosimetry spectrum behind a range shifter (155.3~mm w.e.) for a 150$A$ MeV helium beam. Specific contributions from hydrogen and helium nuclei are shown separately as explained in the figure legend.  Circles represent experimental data~\protect\cite{Tsuda2012} corresponding to the estimated average $^4$He energy of 22$A$~MeV at the entrance to TEPC.}
\end{minipage}
\hspace{0.01\linewidth}
\begin{minipage}[t]{0.5 \linewidth}
\centering
\includegraphics[width=1.0\columnwidth]{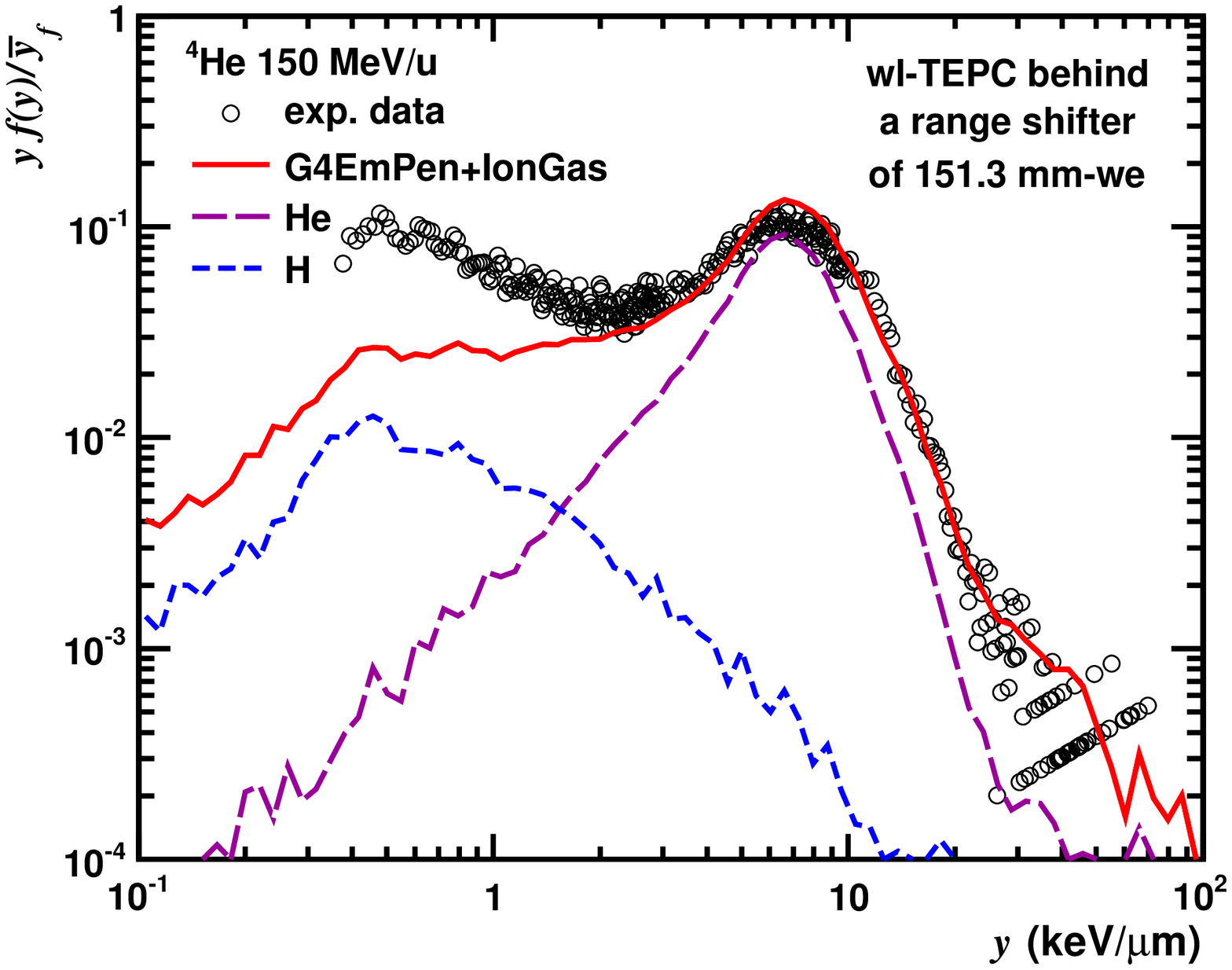}
\caption{\protect\label{fig:spectra_He_32}The same as Fig.~\protect\ref{fig:spectra_He_22} but for a 151.3~mm w.e. range shifter. Circles represent experimental data~\protect\cite{Tsuda2012} corresponding to the estimated average $^4$He energy of 32$A$~MeV at the entrance to TEPC.}
\end{minipage}
\end{figure}

Microdosimetry spectra calculated for helium beams with three physics lists, namely G4EmStd, G4EmPen and G4EmPen+IonGas are presented in Fig.~\ref{fig:spectra_He_17_em}. G4EmStd and G4EmPen give statistically equivalent results which, however, both slightly deviate from the experimental spectrum~\cite{Tsuda2012} at low $y$ and also close to the maximum. Once the models for the gas ionization are involved in calculations in the G4EmPen+IonGas physics list, the agreement with the measured spectrum for $y>$1~keV/$\mu$m is improved, in particular, close to its maximum.  
The shape of microdosimetry spectra for helium beams differs from the one for proton beams. Two distinct peaks are observed in the spectra for $^4$He nuclei. As can be seen from the decomposition of the calculated spectra into contributions of hydrogen and helium nuclei shown in Fig.~\ref{fig:spectra_He_17}, these two prominent peaks are due to hydrogen fragments for events with low $y$ and due to helium nuclei for events with  high $y$. A certain evolution of the position of the helium peak can be observed in Figs.~\ref{fig:spectra_He_17}, ~\ref{fig:spectra_He_22} and~\ref{fig:spectra_He_32} as the energy per nucleon of $^4$He increases. While the projectile energy increases, the helium peak shifts to lower $y$. At the same time there are no noticeable changes in the peak position of hydrogen fragments. This can be explained by the fact that produced hydrogen fragments have an energy spectrum which depends weakly on the beam energy. One can see that the yields of events with $y<1$~keV/$\mu$m in the spectra for the $^4$He beam are underestimated by the MCHIT model, but the reason for this effect may be different as compared to the case of $^1$H beam. The events with $y<1$~keV/$\mu$m  in the spectra calculated for the $^4$He beam are mainly due to secondary hydrogen fragments, see Figs.~\ref{fig:spectra_He_17}, ~\ref{fig:spectra_He_22} and~\ref{fig:spectra_He_32}, and secondary electrons (not shown). This means that the observed effect could also be related to an inaccuracy of the Light Ion Binary Cascade model used to simulate the fragmentation of the $^4$He beam.  

\subsection{Beam of $^7$Li in water}

\begin{figure}[htb]
\centering
\includegraphics[width=1.0\columnwidth]{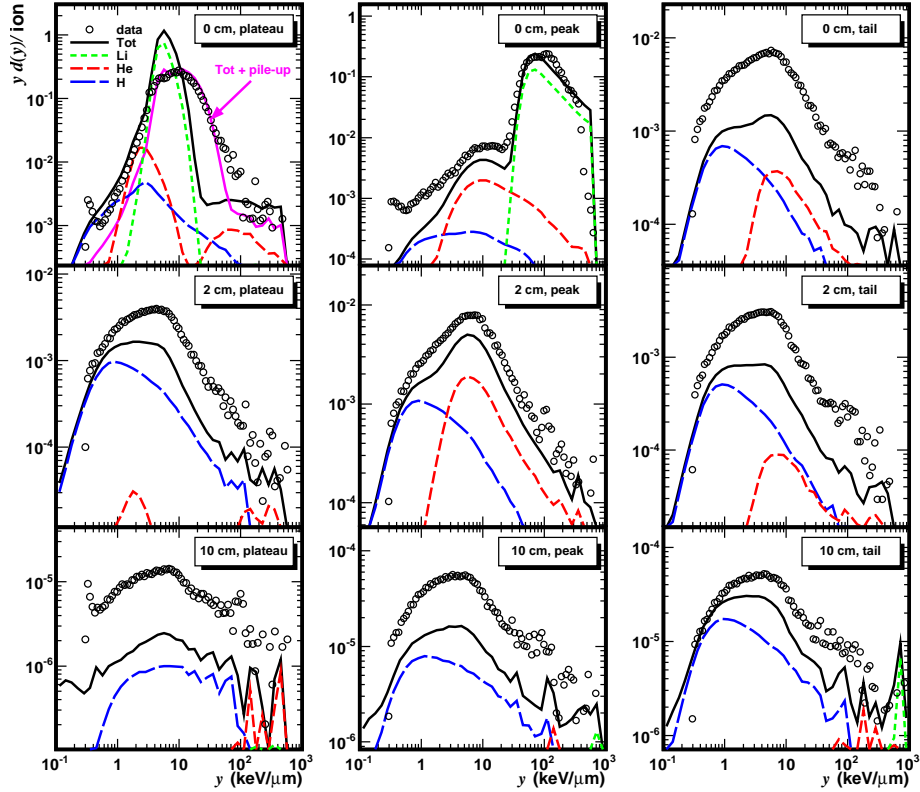}
\caption{\protect\label{fig:spectra_Li7_fragments}Microdosimetry spectra in water phantom irradiated by 185$A$~MeV $^{7}$Li nuclei calculated with MCHIT using G4BIC and G4EmPen+IonGas. Contributions from nuclear fragments of a given charge are shown by various lines as explained in the legend. The distribution at ``0~cm,~plateau'' labelled as ``Tot~+~pile-up'' was obtained with accounting for pile-up events. Circles represent experimental data~\protect\cite{Martino2010}.}
\end{figure}

Calculated and measured~\cite{Martino2010} microdosimetry spectra for lithium beams at nine positions inside a water phantom are shown in Fig.~\ref{fig:spectra_Li7_fragments}. The centre of the TEPC was placed at three positions at the depth of 58.8, 185.8 and 283.8~mm (centers of the gas chamber) corresponding to the plateau, Bragg peak and tail of the depth-dose distribution on the beam axis and also at 2 and 10~cm radial distance from the axis. The microdosimetry spectra on the beam axis typically characterize the radiation field in the center or in front of the target tumour volume during ion therapy.

In addition to the total spectra the contributions from specific nuclei, H, He or Li, are also shown. Prominent peaks of primary $^7$Li nuclei at $y=5$~keV/$\mu$m and $y= 50$~keV/$\mu$m are seen in the $yd(y)$ distributions calculated on the beam axis at the plateau and Bragg peak, respectively.
These peaks are expected to provide the major therapeutic effect. They are superimposed on broader contributions from hydrogen and helium fragments located before and after the main $^7$Li peak. All events registered at the Bragg peak position due to secondary nuclei are characterized by lower $y$ compared to the events due to $^7$Li nuclei. This relation between $y$ of beam nuclei and their fragments also holds at the ``0~cm, plateau'' position. However, in the latter case there exists also a contribution of He nuclei produced in the fragmentation of target nuclei. Since such fragments are slower compared to $^7$Li projectiles and their fragments, such target fragments are responsible for events with high $y$ values, which are seen right to the main peak in the first panel of Fig.~\ref{fig:spectra_Li7_fragments}.

The shapes and positions of the calculated and measured $^7$Li peaks essentially differ at ``0~cm,~plateau''. The calculated peak is sharper as compared to the data and centred at smaller $y$. The $\bar{y}_d$ calculated with MCHIT equals to 7.32~keV/$\mu$m, which is smaller than the value 13.6~keV/$\mu$m estimated from the measured spectrum.  In addition, the average number of events per beam particle in the TEPC placed at ``0~cm,~plateau'' is calculated by MCHIT as $\bar{N}_{c}=0.94$ due to a slight attenuation of the beam after its entrance to the water phantom. According to the normalization of the measured spectra the average number of TEPC events per beam particle in the experiment is only $\bar{N}_{e}=0.48$. All these observations led us to the conclusion that some of the high-y events detected at ``0~cm,~plateau'' were due to two or more $^7$Li nuclei traversing the TEPC within a short time interval and thus appeared  as a single event. Such pile-up events are characterized by an elevated energy deposition to the detector. Since all beam particle histories are modelled by MCHIT independently of each other, pile-up events are impossible in the present simulation. Therefore, providing that the difference between $\bar{N}_{c}$ and $\bar{N}_{e}$ is only due to the pile-up effect, the probability of an event induced by multiple $^7$Li at ``0~cm,~plateau'' can be estimated as    
\begin{equation}
P_{mult} = \frac{\bar{N}_{c}-\bar{N}_{e}}{\bar{N}_{c}} = 0.49 \ .
\end{equation}
One can define the pile-up probability $P_{pu}$ as a probability of a single TEPC event induced by a pair of $^7$Li nuclei. Since multiple events are represented by double, triple etc. coincidence events,
\begin{equation}
P_{mult} = P_{pu} + P^2_{pu} + P^3_{pu} + \dots = \frac{1}{1-P_{pu}} -1 
= \frac{P_{pu}}{1-P_{pu}}  \ ,
\label{eq:pmult}
\end{equation}
resulting in 
\begin{equation}
P_{pu} = \frac{P_{mult}}{1+P_{mult}} = 0.33 \ .
\label{eq:ppileup}
\end{equation}
In order to estimate the contribution of pile-up events to microdosimetry spectra a Monte Carlo method was implemented. The $f(y)$ distribution calculated by MCHIT at ``0~cm,~plateau'' was used to sample independent $y$ events. For each event, the sampled $y$ value is piled-up with the $y$ value of the previous event with a probability $P_{pu}$. The resulting $yd(y)$ distribution is shown in Fig.~\ref{fig:spectra_Li7_fragments} labelled as ``Tot~+~pile-up''. 
As seen, the accounting for pile-up events restores the agreement between calculated and experimental $yd(y)$ distributions. 

Beam nuclei do not reach  TEPC at ``0~cm, tail'' and other six positions, where the spectra are build mainly by secondary H and He fragments and $\delta$-electrons. Also target fragments may eventually contribute with large $y$ events as can be seen at ``10~cm, tail'' but one should keep in mind the poor statistic for such events for TEPC positions at 10~cm away from beam axis. The general shapes of calculated microdosimetry spectra are found to be similar to the shapes of measured spectra. However, the spectra at the tail, 2 and 10~cm away from the beam axis are underestimated by MCHIT. Since at all these positions (excluding ``2~cm, peak'') the spectra are mostly formed by hydrogen-like fragments, this indicates that the yields of proton, deuterons and tritons may be underestimated by the Light Ion Binary Cascade model used to simulate fragmentation of $^7$Li projectiles. For the TEPC position ``2~cm, peak'' this deficiency may also be connected to some underestimation of the yield of helium fragments, as the cascade model neglects the cluster structure of $^7$Li and treats all intra-nuclear nucleons as uncorrelated. It is expected that accounting for the cluster structure of light nuclei would enhance the emission of alpha particles.

\subsection{Beam of $^{12}$C in water}

\begin{figure}[htb]
\centering
\includegraphics[width=1.0\columnwidth]{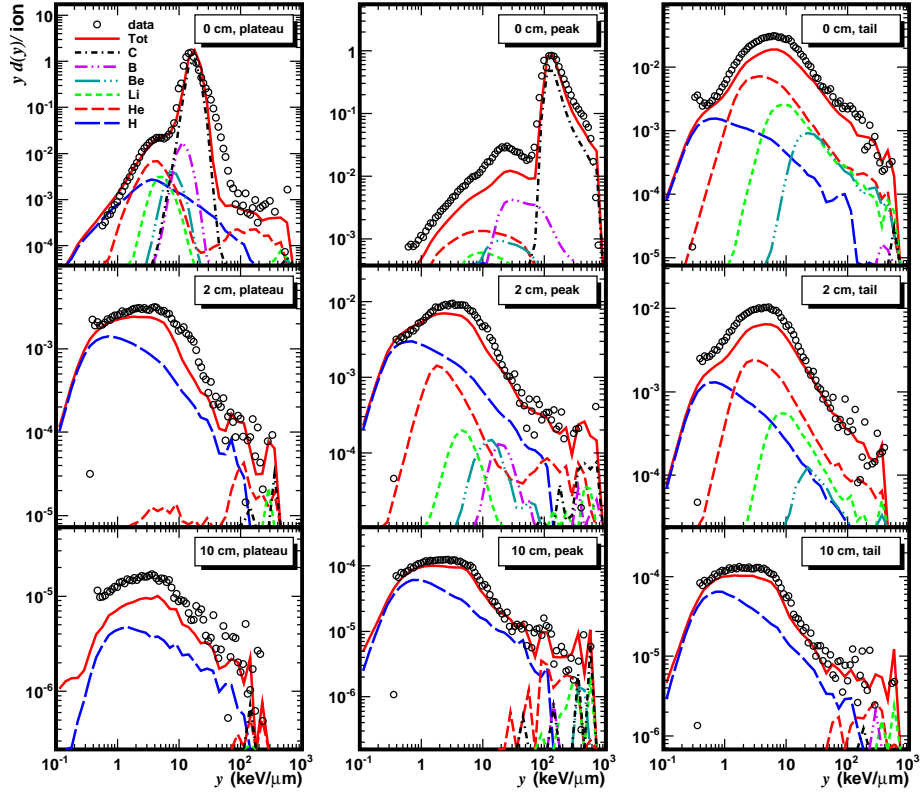}
\caption{\protect\label{fig:spectra_C12_fragments}Microdosimetry spectra in water phantom irradiated by 300$A$~MeV $^{12}$C nuclei calculated with MCHIT using G4QMD and G4EmPen+IonGas. Contributions from nuclear fragments of a given charge are shown by various lines as explained in the legend. Circles represent experimental data~\protect\cite{Martino2010}.
}
\end{figure}

Calculated and measured~\cite{Martino2010} microdosimetry spectra for carbon beams at nine positions inside a water phantom are shown in Fig.~\ref{fig:spectra_C12_fragments}. The exact  TEPC positions used in simulations are given in our previous publication~\cite{Burigo2013}, where further details on the calculational procedure can be found. A shoulder in the spectrum at low $y$ at ``0~cm, plateau'' position is due to the contribution of several projectile fragments. Similar to $^7$Li beam, the events with y$>$100~keV/$\mu$m are due to protons and alphas emitted by target nuclei. Since these target fragments are much slower than the projectile fragments, they provide higher $y$ values at this TEPC position with respect to a sharp peak due to primary $^{12}$C nuclei. Apart from the underestimation of events with 40$<y<$100~keV/$\mu$m, the spectrum measured at the entry to the water phantom on the beam axis is well reproduced by MCHIT. The peak of primary $^{12}$C nuclei also dominates at the TEPC position ``0~cm,~peak'' on the beam axis close to the Bragg peak. There, the agreement with the measured spectrum is in general good with the exception of a slight underestimation of the 
contribution of projectile fragments seen at lower $y$ before the main peak.    

The spectra at ``0~cm,~tail'', ``2~cm,~peak'' and ``2~cm,~tail'' are built from overlapping contributions from various projectile fragments: H, He, Li and Be. Among these three positions a noticeable contribution from boron nuclei is predicted only at ``2~cm,~peak''. As follows from the calculations, the maxima of the contributions from H, He, Li, Be and B are ordered according to $Z^2$ of the corresponding nuclei: more heavy fragments contribute with larger $y$. The contributions from He nuclei are remarkable. The spectra at these three positions are also underestimated, as in the case of $^7$Li beam. The quantitative agreement between calculations and measurements for $^{12}$C beam is much better compared to $^7$Li beam, but still the deviations of the calculated spectra from measured ones can be attributed to the underestimation of He fragments.  The other four spectra, namely, at ``2~cm,~plateau'', ``10~cm,~plateau'', ``10~cm,~peak'' and ``10~cm,~tail'' are mostly composed from the contributions of hydrogen nuclei produced in fragmentation of $^{12}$C. They are also slightly underestimated by MCHIT.  Some traces of the contributions from fragments of the target nuclei are also seen at these four positions far from the $^{12}$C beam. Such target fragments are scarcely produced in nuclear reactions induced by energetic secondary protons and neutrons.

\subsection{RBE and biological dose profiles for $^1$H, $^4$He, $^7$Li and $^{12}$C beams}

\begin{figure}[htb!]
\centering
\includegraphics[width=0.8\columnwidth]{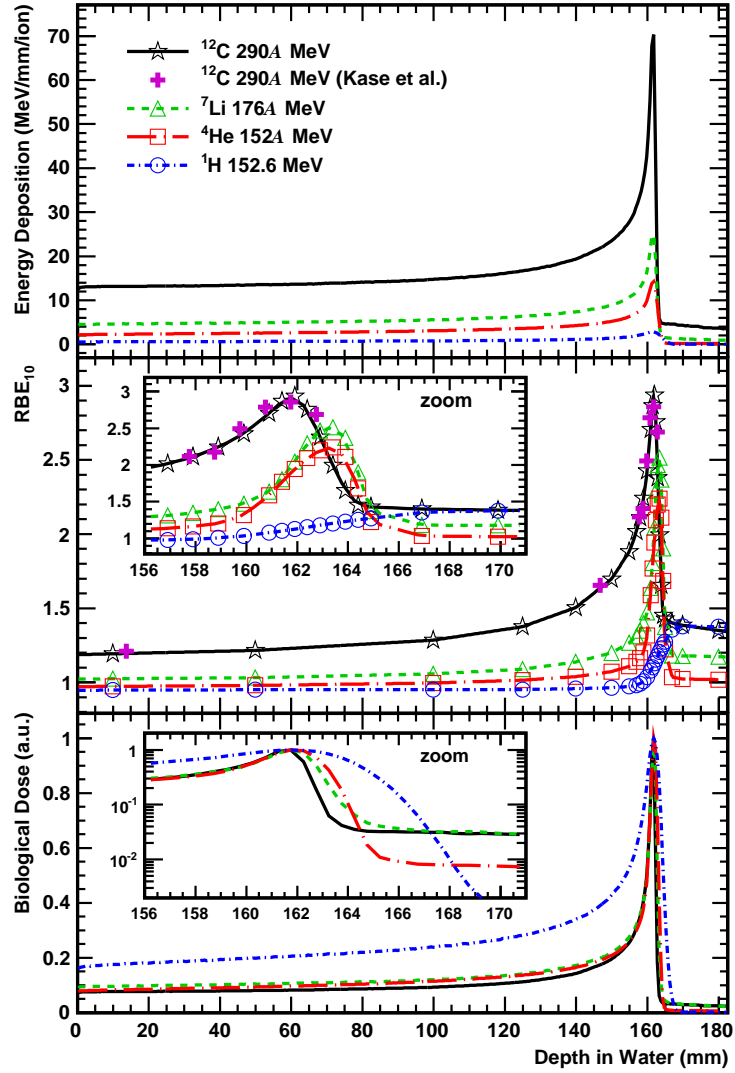}
\caption{Calculated energy deposition profile per ion beam in water for $^1$H, $^4$He, $^7$Li and $^{12}$C beams with Bragg peak at 161.6~mm (top panel), estimated RBE$_{10}$ for HSG cells (middle panel) and biological dose (bottom panel). The cross symbols (\ding{58}) show the RBE$_{10}$ calculated from experimental values of $y^*$~\protect\cite{Kase2006} while other symbols present calculated results by MCHIT with modified MK model. The biological doses for all ions were rescaled by respective values at the Bragg peak.}
\label{fig:RBE}
\end{figure}

Using the calculated microdosimetry spectra we may now estimate the biological effectiveness of $^1$H, $^4$He, $^7$Li and $^{12}$C nuclei. For the sake of comparison we have chosen their beam energies such that they lead to similar ranges in water. The microdosimetry spectra were calculated at several positions along the beams axes. The corresponding energy deposition profiles (bin size of 0.1~mm) are shown in logarithmic and linear scales in the top panels of Figs.~\ref{fig:depth_profiles} and~\ref{fig:RBE}, respectively.

As demonstrated in the previous sections, the microdosimetric spectra depend strongly on depth and radial distance from the beam axis. Such variation in the energy deposition pattern leads to very different biological effects. The values of RBE$_{10}$ estimated by means of MCHIT coupled with the modified MK model on the beam axis as a function of depth are presented in the middle panel of Fig.~\ref{fig:RBE} for the four ion beams.  Experimental RBE$_{10}$ for carbon beam estimated from experimental data~\cite{Kase2006} are plotted in the same figure for comparison. The well-known increase of the RBE for carbon ions on their way from the plateau region to the Bragg peak seen in the experimental data is well reproduced by MCHIT+MK model. Helium and lithium ions also show favourable RBE profiles characterized by even lower values at the plateau region, with much steeper increase close to the Bragg peak and lower RBE values in the tail. The maximum RBE$_{10}$ values for helium, lithium and carbon ions found around the Bragg peak are 2.2, 2.5 and 2.9, respectively. At the plateau region the values are 1.0, 1.0 and 1.2, respectively. For protons, the model predicts the RBE$_{10}$ value slightly below 1 at the entrance to the phantom and a smooth increase to 1.2 at the Bragg peak. A further increase of RBE$_{10}$ after the peak may be related to slow secondary neutrons produced in nuclear reactions. Similar calculations were performed with the TEPC placed 2 cm away from the beam axis at the depth of the Bragg peak. At this point RBE$_{10}$ of 1.1$\pm$0.1 was estimated from calculated microdosimetry spectra for all four ions.

The biological dose at the considered TEPC positions can now be estimated as the product of RBE and the physical dose. The results for the biological dose for different ions are shown in the bottom panel of Fig.~\ref{fig:RBE}. The curves were rescaled in order to yield the same dose at the Bragg peak. The biological dose profiles for the helium, lithium and carbon beams were found to be similar to each other. All three ions demonstrate a high ratio of the dose at the Bragg peak relative to the dose at the plateau which helps to spare healthy tissues traversed by the beam before reaching a tumour. The biological dose values at the tail of the lithium and carbon beams are found to be very similar, while the helium beam delivers a smaller dose to the tail region due to a reduced fragmentation rate. These results show that helium and lithium beams are also promising options in addition to a well-established carbon-ion cancer therapy.

\begin{figure}[htb]
\centering
\includegraphics[width=0.8\columnwidth]{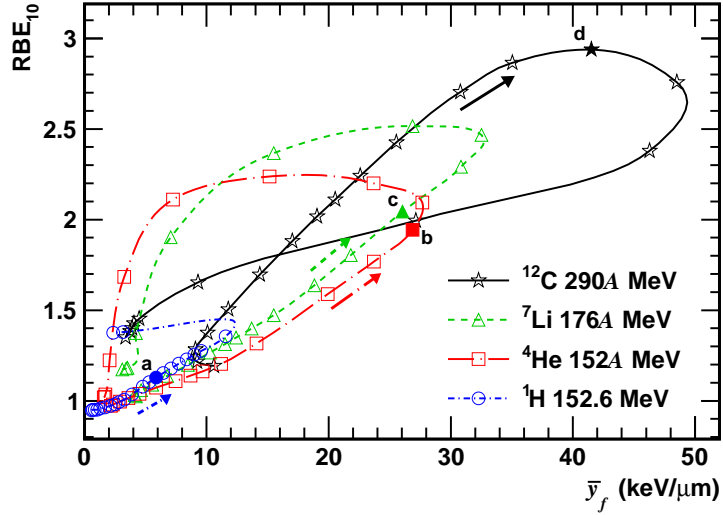}
\caption{\label{fig:RBE_yf} Calculated correlation between RBE$_{10}$ and $\bar{y}_f$ for $^1$H, $^4$He, $^7$Li and $^{12}$C beams in water. The lines connect the values obtained for TEPCs sequentially placed along the beam axes while arrows indicate the direction for increasing depth. Full symbols indicated by letters `a', `b', `c' and `d' correspond to the values at the Bragg peak for protons, helium, lithium and carbon ions, respectively.}
\end{figure}

Despite the fact that the depth-dose and RBE$_{10}$ profiles shown in Fig.~\ref{fig:RBE} look similar in shape, the relation between $\bar{y}_f$ representing LET and RBE$_{10}$ is not trivial. The correlated pairs of values $\bar{y}_f$ and RBE$_{10}$ are shown in Fig.~\ref{fig:RBE_yf} for all four considered beams. The curves demonstrate a monotonic increase of RBE$_{10}$ with the rise of $\bar{y}_f$ before the Bragg peak. The correlations for $^4$He and $^7$Li are similar to each other, but they differ from the case of $^{12}$C. In the case of the carbon beam, the increase is more pronounced but RBE$_{10}$ reaches a maximum at the Bragg peak and then slightly drops for increasing $\bar{y}_f$ values. This behaviour shows a kind of saturation effect. Such effect is not observed for $^4$He and $^7$Li. Particularly, one can see that for these two ions RBE$_{10}$ continues to rise even when $\bar{y}_f$ values start decreasing in the tail region. Another feature of the relation between $\bar{y}_f$ and RBE$_{10}$ is that for a given beam similar $\bar{y}_f$ values correspond to quite different RBE$_{10}$ at the plateau and tail of the depth-dose distribution. Primary beam particles and secondary fragments which dominate, respectively, in these two regions provide rather similar $\bar{y}_f$, but their RBE$_{10}$ differ significantly. These results confirm that $\bar{y}_f$ (LET) solely is not sufficient to characterize the biological effects of various beams.

\section{Conclusion}\label{sec:conclusions}

Our analysis of the microdosimetry spectra for light nuclei lead us to the following conclusions:

\begin{itemize}
\item The microdosimetry spectra of protons calculated for the macroscopic-size TEPC filled with dilute gas agree well with the microdosimetry spectra calculated for the equivalent microscopic volume of water with G4DNA physics list. In this way the basic assumption of the microdosimetry technique is fully validated by Monte Carlo simulations with MCHIT. 

\item Contributions of primary beam nuclei and secondary fragments to the microdosimetry spectra can be realistically evaluated by Monte Carlo simulations with MCHIT. 

\item A proper modelling of nuclear fragmentation reactions is crucial for describing microdosimetry spectra both on the beam axis and far from the beam. Further improvements of nuclear fragmentation models, e.g. with respect to production of $^4$He nuclei, could improve the description of the microdosimetry data and, therefore, provide a better understanding of the radiation effects from the considered therapeutic beams.

\item The MCHIT model is able to describe reasonably well the microdosimetry spectra for hydrogen and helium beams in water. Microdosimetry spectra for lithium beams are well described on the beam axis from the entrance down to the Bragg peak while the spectra are underestimated far from the primary beam. It was demonstrated that in the case of TEPC directly irradiated by a $^7$Li beam, the agreement with the experimental data can be significantly improved by taking into account the pile-up effect. This is specific to the experimental data of~\cite{Martino2010}. It is expected that similar measurements, but with lower beam current, will be not distorted by overlapping events. The spectra for carbon beam are generally well described, despite of some underestimations at positions far from the beam axis and also in the tail region. 

\item MCHIT coupled with the modified MK model allowed us to estimate the RBE for proton, helium, lithium and carbon ions at several positions in a water phantom. The models predict favourable biological dose-depth profiles for helium and lithium beams similar to the one for carbon beam. This result suggests that helium and lithium beams could also be used for cancer therapy.

\item The correlations between RBE$_{10}$ and $\bar{y}_f$ for proton, helium, lithium and carbon ions were studied along the beam axis. Such a correlation for carbon beam reveals the saturation effect. It is found that $\bar{y}_f$ (representing LET) may be similar at the plateau and tail regions, but still lead to very different biological effects. 

\item Finally we want to emphasize that the main conclusion of this work is that helium and lithium beams should be considered as rather favourable options for cancer therapy. They have reduced fragmentation cross sections compared to the $^{12}$C beam that makes them preferable for a deeply-seated tumour. On the other hand, they have a reduced lateral scattering compared to the proton beam. At the same time the biological effectiveness of these beams is only slightly lower than that of $^{12}$C beam.
\end{itemize}

\section*{Acknowledgements}
This work was carried out within the framework of NanoBIC-NanoL project.
L.B. is grateful to the Beilstein Institute for the financial support. This work was also partially supported by HIC for FAIR within the Hessian LOEWE-Initiative.
We wish to thank G.~Martino and S.~Tsuda for sending us their tables of experimental data.
Our calculations were performed at the Center for Scientific Computing (CSC) of the
Goethe University Frankfurt. We are grateful to the CSC staff for support.

\bibliographystyle{unsrt}

\end{document}